\documentclass[11pt, epsf]{article}

\usepackage{hyperref}
\usepackage[bf]{caption}
\usepackage{epsfig}
\usepackage{float}
\usepackage{graphicx}
\usepackage{subfig}
\def\ga{\mathrel{\raise.3ex\hbox{$>$\kern-.75em\lower1ex\hbox{$\sim$}}}}
\def\la{\mathrel{\raise.3ex\hbox{$<$\kern-.75em\lower1ex\hbox{$\sim$}}}}

\def\I_M{{I_{\scriptscriptstyle M\times M}}}

\begin{document}

\thispagestyle{empty}

\vskip 2cm \centerline{ \Large \bf  Anisotropic Power Law Inflation 
from Rolling Tachyons} 

\vskip .2cm

\vskip 1.2cm

\centerline{ \bf Samrat Bhowmick and 
Sudipta Mukherji\footnote{Electronic address: 
samrat, mukherji@iopb.res.in}}
\vskip 10mm \centerline{ \it Institute of Physics, 
Bhubaneswar-751 005, India} 
\vskip 1.2cm
\vskip 1.2cm
\centerline{\bf Abstract}
\noindent

We provide an explicit solution representing an anisotropic power law 
inflation
within the framework of rolling tachyon model. This is generated 
by allowing a non-minimal coupling between the tachyon and the world-volume
gauge field on non-BPS $D3$ brane. We also show that this solution is perturbatively 
stable. 

\newpage
\setcounter{footnote}{0}
\noindent

If the origin of the anomalies in the WMAP data, namely, the observed 
power 
asymmetry \cite{Hansen:2008ym,Eriksen:2003db}, the alignment of low-$l$ 
CMB multipoles 
\cite{Tegmark:2003ve,Bielewicz:2004en,Land:2005ad,Copi:2005ff}, 
happens to be 
primordial, it will imply statistical anisotropy of the primordial 
curvature perturbation. How can one generate such anisotropy with-in
the simplest scalar field models of inflation? Scalar field does not 
prefer any particular direction and hence it is hard to imagine 
that anisotropies can be created by scalars. This motivated 
many researchers to look into models of inflation with vector fields.
In some of these models such as \cite{Ackerman:2007nb}, vector fields do 
not 
satisfy dominant energy 
condition and thus evade the cosmic no-hair 
conjecture\cite{Wald:1983ky}. Though, these 
models provide examples of anisotropic inflation, they are plagued with 
instabilities \cite{Dulaney:2008ph, Himmetoglu:2008hx}.

Recently, another model of anisotropic inflation was proposed in 
\cite{Watanabe:2009ct} and corresponding curvature perturbation power 
spectra was studied in \cite{Dulaney:2010sq,Watanabe:2010fh}.
For further studies in this model, see \cite{Watanabe:2010bu,Murata:2011wv}.
This model includes a non-minimal coupling between the gauge field and 
the inflaton. Such models typically lead to power law inflation rather 
than the deSitter ones. In particular, consider the scenario in 
\cite{Kanno:2010nr}. It starts with an action of the form
\begin{equation}
S = \int d^4x {\sqrt{-g}}\Big[ \frac{M_p^2}{2} R - \frac{1}{2} 
\partial_\mu\phi \partial^\mu \phi - V(\phi) - \frac{1}{4} f^2(\phi) 
F_{\mu\nu}F^{\mu\nu} \Big],
\end{equation}
with $V(\phi) = V_0 e^{\frac{\lambda \phi}{M_p}}, f(\phi) = f_0 
e^{\frac{\rho\phi}{M_p}}$. Here $M_p$ is the four dimensional Planck 
mass and $\lambda, \rho$ are two constants. Taking $A_\mu = (0, A_x(t), 
0, 0)$ and $\phi = \phi(t)$, it is possible to consistently solve the 
Einstein, scalar and vector equations for metric of the form
\begin{equation}
ds^2 = - dt^2 + t^p ~[ ~t^{-q} dx^2 + t^{q/2} (dy^2 + dx^2) ~],
\end{equation}
for some $p >>1$ and $q > 0$. Further, an analysis of the dynamical 
system shows that the above power law solution in in fact an attractor
for a large range of parameters\footnote{For other discussions on 
anisotropic inflations within this scenario, see for example 
\cite{Emami:2010rm, Dimopoulos:2010xq, Karciauskas:2011fp}}. 

Motivated by these developments, in this paper, we consider anisotropic 
inflation in models driven by rolling tachyons. A while back, in 
\cite{Sen:2002nu,Sen:2002in}, classical time dependent 
solution describing the decay process of an unstable D-brane in 
the open string theory was constructed. During this process, the tachyon 
field on the brane rolls down to the minimum of the potential. In these 
works, it was also pointed out that this rolling tachyon might have 
cosmological significance. Since then there are plethora of activities 
starting with \cite{Gibbons:2002md, Mukohyama:2002cn, 
Feinstein:2002aj, Padmanabhan:2002cp, Jones:2002cv, Choudhury:2002xu}. 
There are several 
reasons for that.
Firstly, this model directly arises from 
string theory. Secondly, it promises potential applications in the 
inflationary scenario of our universe.
However, rolling tachyon  models also have their own problems. This is  
discussed, for example, in \cite{Kofman:2002rh}. There, it was 
argued that tachyonic inflation, in general, can not result in a 
universe  
which is reasonably close to our observed one. More precisely, for the
tachyonic potential $V(T)$, that is expected to arise from string 
theory,
there is an incompatibility between the slow role condition and the COBE 
normalisation of fluctuations. There are also few other problems 
associated with these models. It was however noted that some of these 
issues get cured if one works with a model of assisted tachyonic 
inflation\footnote{See \cite{Liddle:1998jc,Mazumdar:2001mm} for models of 
assisted inflation.} . Here, one considers $N$ 
non-BPS D3 brane. In this system, there
are two different kinds of open strings - one that stretches between
two different branes and the others whose both ends are on the same 
brane.
If the distance between the branes are larger than the string scale,
one can neglect the first set of strings. Consequently, the theory 
reduces to a system of quite non-interacting tachyons, one on every brane. If 
number of tachyons is of the order of $10^{11}$, it seems that there 
is a compatibility between slow role 
condition and the COBE normalisation of fluctuations. However some of 
the 
other problems still remain \cite{Piao:2002vf}. 

The effective field 
theory Lagrangian on the D3 brane, in general, has a Born-Infeld
form and is given by
\begin{equation} 
{\cal L} = \sqrt{-g} \; L = - V(T) \sqrt{-g} \;
{\sqrt{\det[\delta^\mu{}_\nu+ 
h(T) F^\mu{}_\nu + \partial^\mu T \partial_\nu T] }} \;.
\label{bi}
\end{equation}
However, for non-interacting $N$-tachyon ($T$) assisted inflation, one 
works
with the Lagrangian
\begin{equation}
{\cal L} = \sqrt{-g} \;L = - N V(T) \sqrt{-g} \;
{\sqrt{\det[\delta^\mu{}_\nu + 
h(T) F^\mu{}_\nu + \partial^\mu T \partial_\nu T] }} \;.
\label{nbi}
\end{equation}
Our humble aim in this paper is to show that besides 
usual inflationary expansion \cite{Feinstein:2002aj, 
Padmanabhan:2002cp}, 
for a suitable choice of $V(T)$\footnote{On general grounds, it is 
expected that the potential has a maximum near $T = 0$ and decays off
exponentially at large $T$. We will work here with 
\begin{equation}
V(T) = \frac{V_0}{T^2}, ~~{\rm with}~~V_0 > 0.
\label{pot}
\end{equation}
In spite of the divergence at $T =0$, $V(T)$ mimics quite closely
the expected tachyon potential with an advantage that it provides
exact solution of a power law inflation.}, this Lagrangian also provides us
with an exact solution for anisotropic inflation. This happens  if we 
allow for a suitable tachyon-gauge field coupling through the function 
$h(T)$ in (\ref{nbi}). We further show that this solution is perturbatively 
stable. 

We parametrise our metric as in \cite{Watanabe:2009ct}:
\begin{equation}
ds^2 = -dt^2 + e^{ 2 \alpha(t) - 4\sigma(t)} dx^2 + 
e^{2 \alpha(t) + 2 \sigma(t)} (dy^2 + dz^2).
\end{equation}
It is easy to derive the equations of motion for $g_{\mu\nu}, A_\mu$
and $T$ that follow from the coupled action
\begin{equation}
S = \int d^4x {\sqrt{-g}}\Big[\frac{M_p^2}{2} R + L].
\end{equation}
Here $R$ is the usual four dimensional Ricci scalar.
These are\footnote{In general, with the full contribution from the
determinant in $L$ in the equations of motion, it is hard to solve
the differential equations. We have approximated the determinant 
by expanding it in gauge field and scalar fields while
keeping only the first terms both gauge field and scalar fields. 
We point out that in our way of writing, $h(T)$ has an 
overall $\alpha^\prime$ 
dependence. We will discuss the range of the parameters later.}
\begin{eqnarray}
&& (\partial_t \alpha)^2 = (\partial_t \sigma)^2 + \frac{N}{3} 
\Big[ \Big(1 + \frac{1}{2} (\partial_t T)^2\Big) V + \frac{p^2}{2 U} 
e^{-4 \alpha - 4 \sigma}\Big], \nonumber\\
\nonumber\\
&& \partial_t^2 \sigma = - 3 \partial_t \alpha ~\partial_t \sigma + 
\frac{N 
p^2}{3 U} e^{-4 \alpha - 4 \sigma}, \nonumber\\
\nonumber\\
&& \partial_t^2 \alpha = -3 (\partial_t \alpha)^2 + N V 
+ \frac{N p^2}{6 U} e^{-4 \alpha - 4 \sigma}, \nonumber\\
\nonumber\\
&& V~\partial_t T \Big[ \partial_t^2 T +  3 \partial_t \alpha 
~\partial_t 
T + \frac{1}{V} \frac{dV}{dT} \Big(1 + \frac{1}{2} (\partial_tT)^2\Big)\Big]
- \frac{1}{2}\frac{p^2  \partial_t U}{U^2} e^{-4 \alpha - 4 \sigma} = 0.
\label{eom}
\end{eqnarray}
In these equations, we have defined $U = h^2 V$. Furthermore, 
we took $A_\mu (t) = (0, A_x(t), 0, 0)$. Using 
this, we integrated the gauge field equation of motion as
\begin{equation}
\partial_t A_x (t) = \frac{p}{U} e^{- \alpha - 4\sigma }
\label{gfe}
\end{equation}
and used this in (\ref{eom}). The parameter $p$ in (\ref{gfe}) arises as 
an integration constant.

Out of the four equations 
in (\ref{eom}), only three equations are however independent. To see 
this, we differentiate the first and use the
last equation to get 
\begin{equation}
2 \partial_t \alpha ~\partial_t^2 \alpha = 2 \partial_t \sigma 
~\partial_t^2 \sigma - N V \partial_t \alpha ~\partial_t^2 T
- \frac{2 N p^2}{3 U}  e^{-4 \alpha - 4 \sigma} \Big(\partial_t \alpha + 
\partial_t \sigma\Big).
\label{inter}
\end{equation}
Further using the first and second equation of (\ref{eom}), we can 
reduce (\ref{inter}) to the third of (\ref{eom}). Therefore, for now on,
we will only consider the first, second and the fourth as independent 
equations of motion.

For the potential (\ref{pot}), there is an exact solution for the metric
and other fields. These are given by
\begin{equation}
\label{noaniso}
\alpha = \frac{NV_0}{2} {\rm log}~t, ~~\sigma = 0, ~~T = \frac{2 
t}{\sqrt{3 N V_0 
-2}}, ~~p = 0, ~~h(T) = 0.
\end{equation}
Clearly, this leads to an isotropic inflationary power law expansion.
The inflationary slow-roll parameters can be adjusted by tuning $N V_0$.

We  now explicitly construct another simple solution, for the same 
potential, 
which provides us with an anisotropic accelerated expansion of the 
universe\footnote{For other discussions on anisotropic inflation, see 
\cite{Moniz:2010cm}. The solution we construct here is new.}. To 
this end, let us define
\begin{equation}
\alpha = \alpha_0 ~{\rm log}~t, ~~ \sigma = \sigma_0 ~{\rm log}~t, ~~h = 
t^\delta, ~~T = T_0 t.
\label{bc}
\end{equation}
We will now substitute these in (\ref{eom}) and solve for 
$\alpha_0, \sigma_0, \delta$ and $T_0$.
Using (\ref{bc}) in (\ref{eom}), we get the following restrictions
on the parameters
\begin{eqnarray}
&& 2  - \delta -2 \alpha_0 - 2 \sigma_0 = 0, \nonumber\\
\nonumber\\
&&\frac{N p^2 T_0^2}{V_0} + N V_0~\Big(1 + \frac{2}{T_0^2}\Big) - 6 
\alpha_0^2 
+ 6 \sigma_0^2 = 0, \nonumber\\
\nonumber\\
&& 3 \Big( 3 \alpha_0 -1\Big) \sigma_0 - \frac{N p^2 T_0^2}{V_0} = 0, 
\nonumber\\
\nonumber\\
&& p^2 ~T_0^4 ( 1 - \delta  ) + T_0^2 ~V_0^2 ( 3 
\alpha_0 -1) - 2 V_0^2 = 0.
\label{rest}
\end{eqnarray}
These equations can be further simplified by a replacement $p \rightarrow 
p V_0$. 
\begin{eqnarray}
&& 2  - \delta -2 \alpha_0 - 2 \sigma_0 = 0, \nonumber\\
\nonumber\\
&& NV_0\left[p^2 T_0^2  + 1 + \frac{2}{T_0^2}\right] - 6
\alpha_0^2
+ 6 \sigma_0^2 = 0, \nonumber\\
\nonumber\\
&& 3 \Big( 3 \alpha_0 -1\Big) \sigma_0 - N V_0 p^2 T_0^2 = 0,
\nonumber\\
\nonumber\\
&& p^2 ~T_0^4 ( 1 - \delta ) + T_0^2  ( 3
\alpha_0 -1) - 2 = 0.
\label{restric}
\end{eqnarray}

On substituting $\sigma_0$ from the third equation into the
second, we get
\begin{equation}
2 N^2 V_0^2 p^4 T_0^6 + 3 (1 - 3 \alpha_0)^2 [(2 + T_0^2 + p^2 T_0^4)NV_0 - 6 \alpha_0^2 T_0^2 ]= 0.
\label{one1}
\end{equation}
Further using the first and the third into the fourth of 
(\ref{restric}), we have
\begin{equation}
2 N V_0 p^4 T_0^6 + 3 ( 3 \alpha_0 - 1) [T_0^2 \{p^2 T_0^2 
(2 \alpha_0 -1) + 3 \alpha_0 -1\} - 2]=0.
\label{two2}
\end{equation}
Now, given the parameters $N V_0$ and $p$, (\ref{one1}) and 
(\ref{two2}) are two equations for two unknowns
$\alpha_0$ and $T_0$. Though
the above equations can be solved exactly for $\alpha_0$ and $T_0$, 
the solutions involve rather large expressions and non-illuminating.
We, instead,  prefer to plot the solutions.  

\begin{figure}[H]
 \centering

 \subfloat[]{ 
 \includegraphics[width=0.3\textwidth]{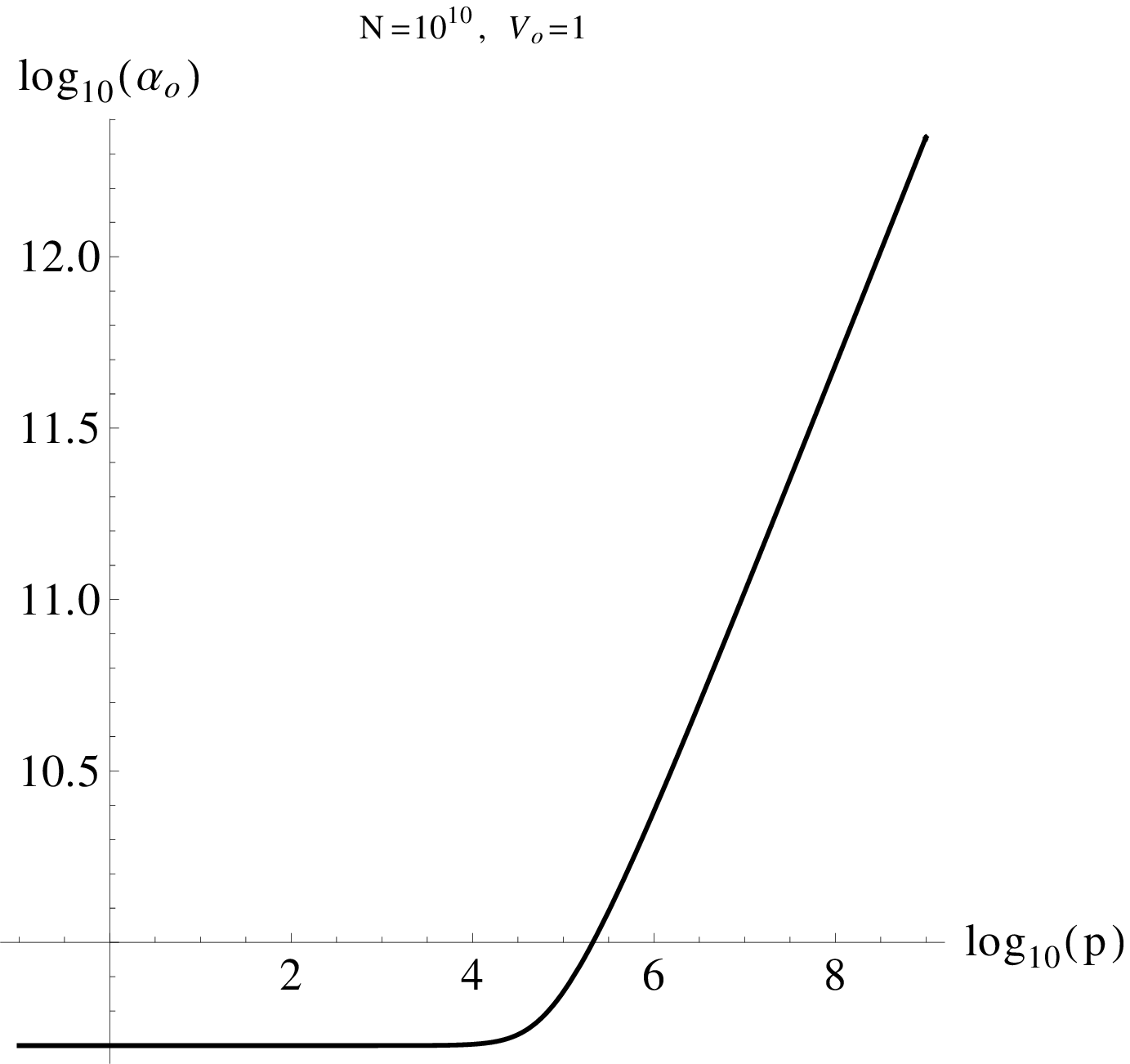}} ~~~~~~~
 \subfloat[]{
 \includegraphics[width=0.3\textwidth]{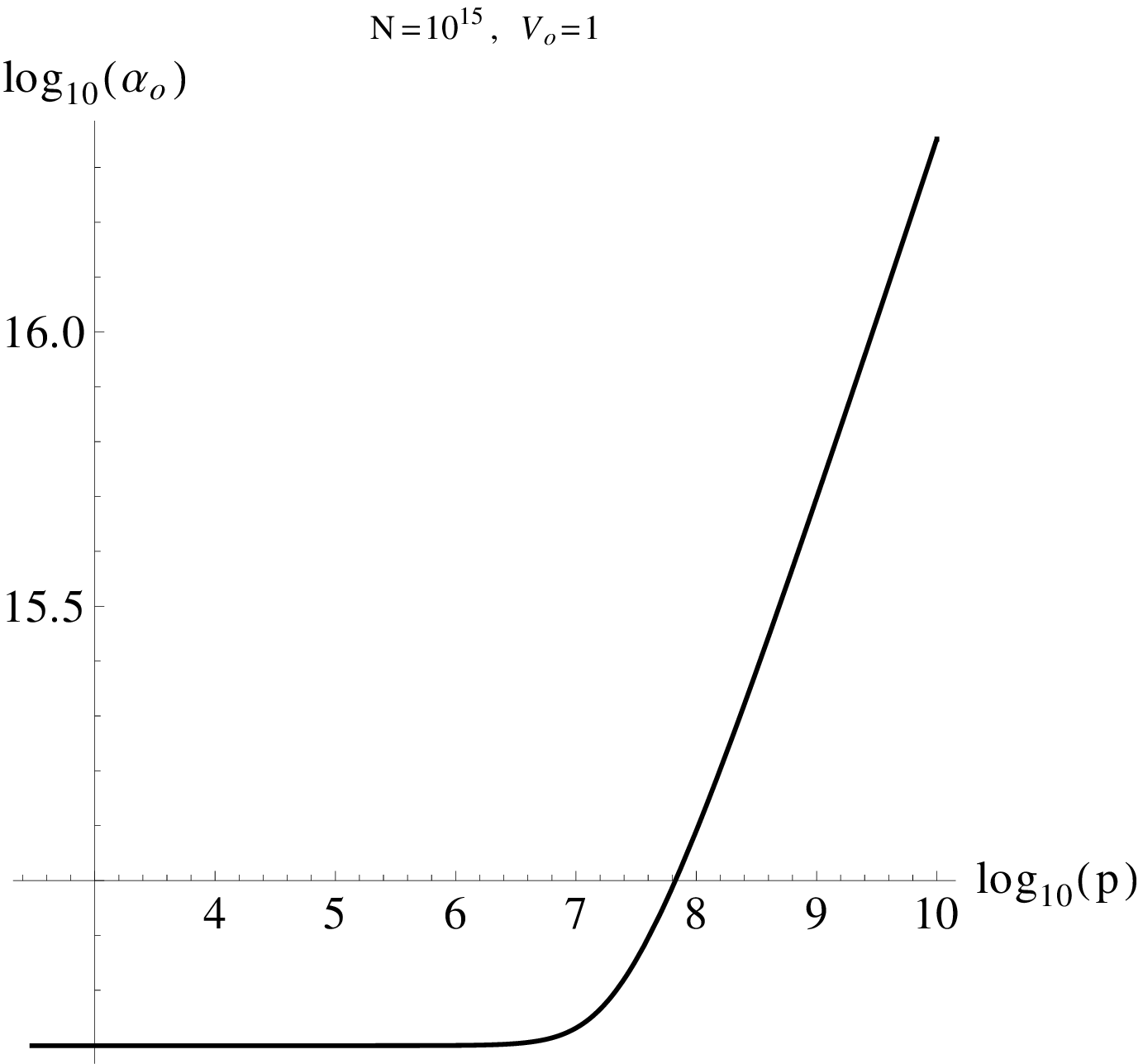}}

 \subfloat[]{
 \includegraphics[width=0.3\textwidth]{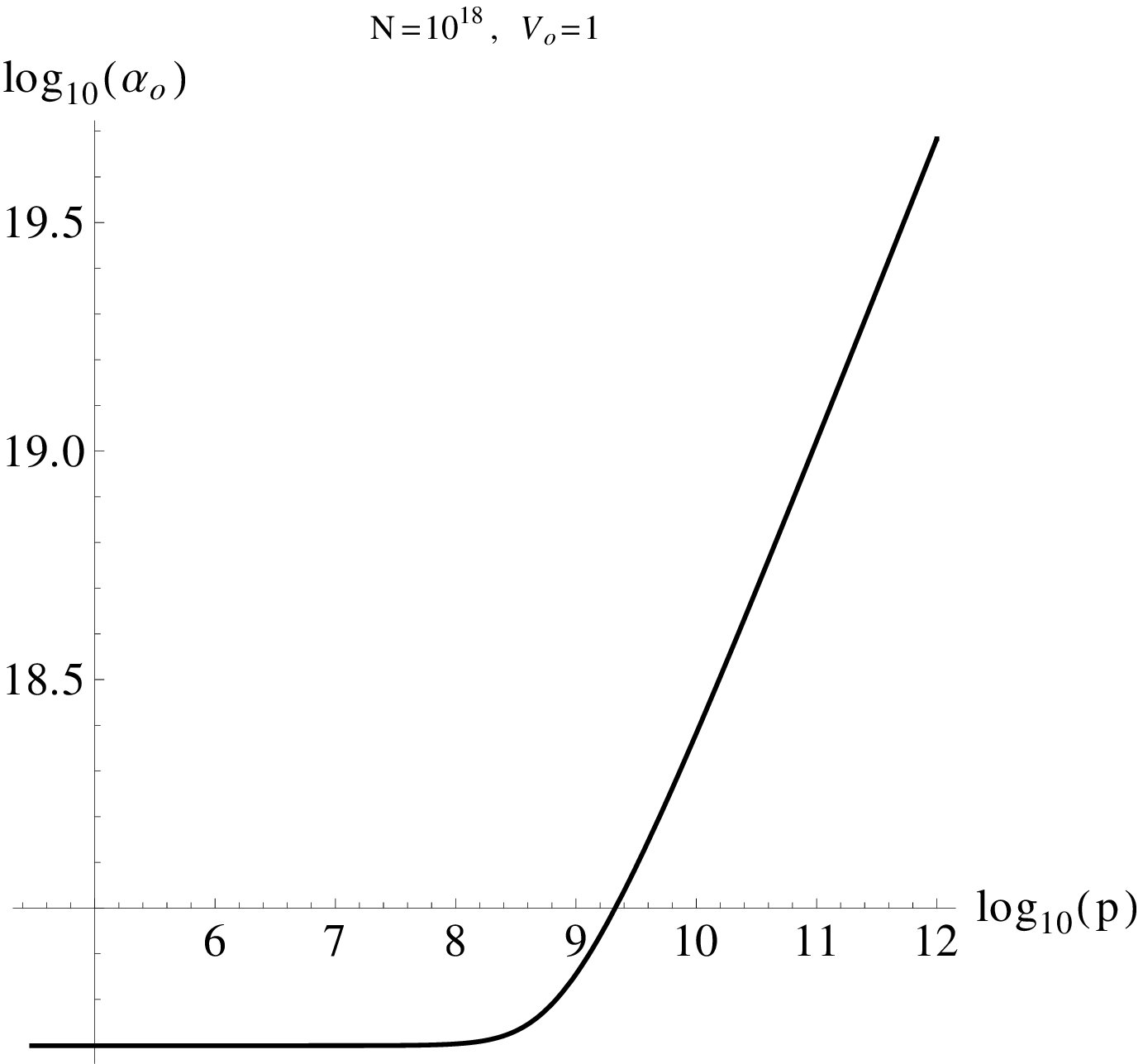}}

 \caption{ 
{\em In these figures, $\log(\alpha_0)$ vs $\log(p)$ 
has been ploted with different values of $N$. $N=10^{10}$, $N=10^{15}$ and
$N=10^{18}$ in {\bf \em figure} ({\bf a}), ({\bf b}) and ({\bf c}) 
respectively. We have fixed $V_0 =1$.
}
}
\label{fig:alpha}
\end{figure}

\begin{figure}[H]
 \centering

 \subfloat[]{
 \includegraphics[width=0.3\textwidth]{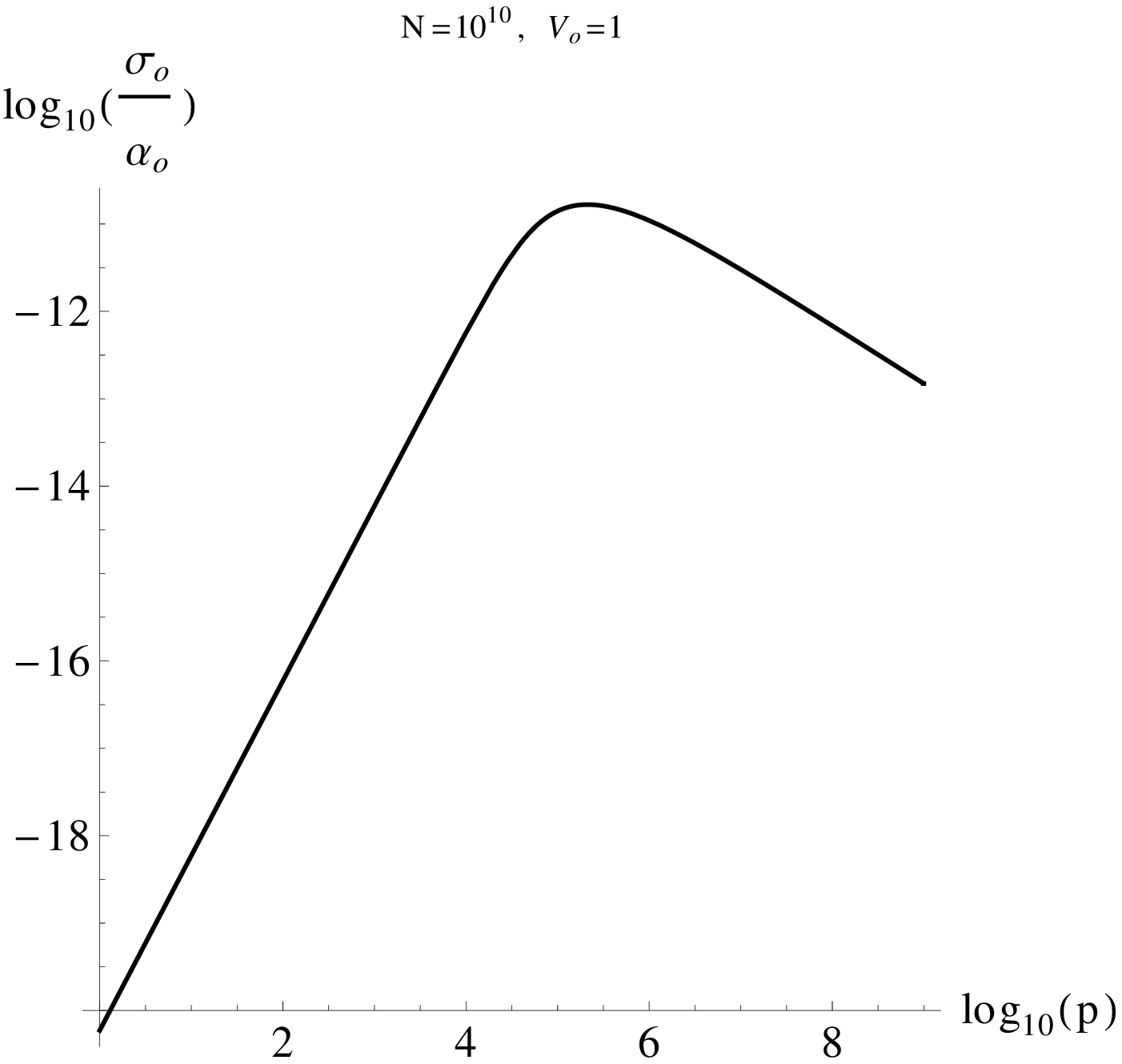}} ~~~~~~~
 \subfloat[]{
 \includegraphics[width=0.3\textwidth]{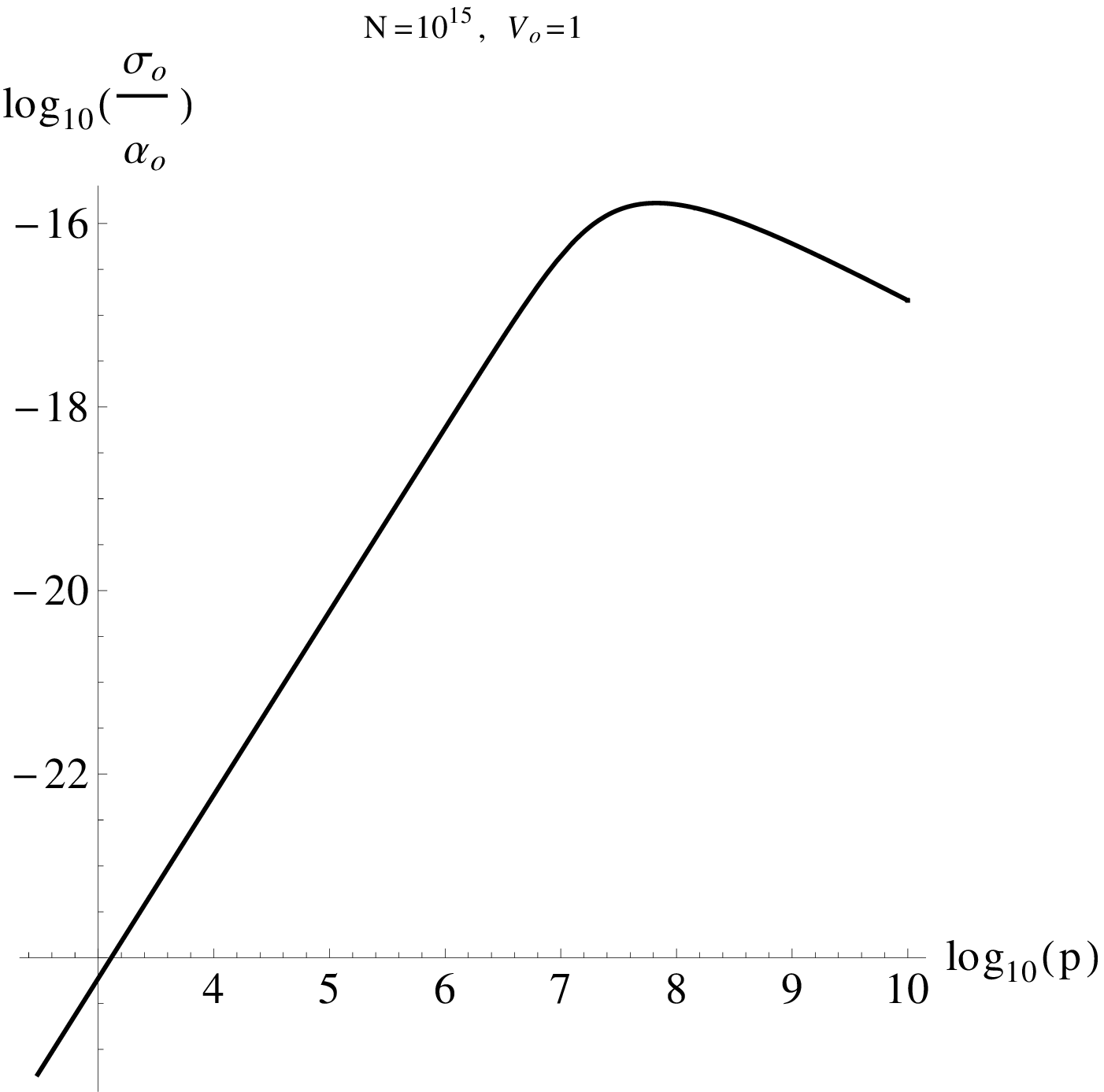}}

 \subfloat[]{
 \includegraphics[width=0.3\textwidth]{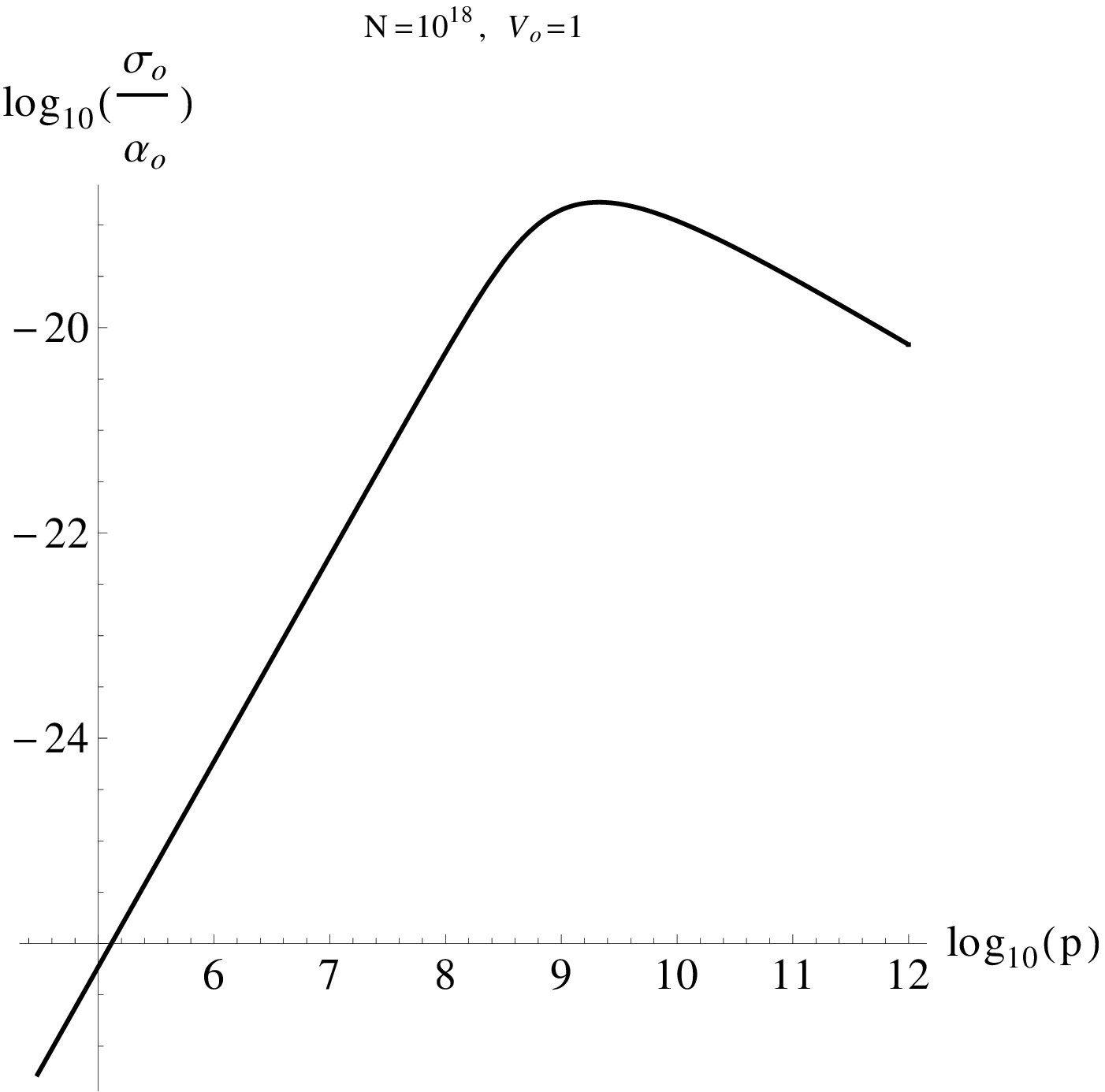}}

 \caption{ 
{\em In all the figures, $\log(\frac{\sigma_0}{\alpha_0})$
vs $\log(p)$ 
has been ploted with different values of $N$. $N=10^{10}$, $N=10^{15}$ and
$N=10^{18}$ in {\bf \em figure} ({\bf a}), ({\bf b}) and ({\bf c}) 
respectively. We have fixed $V_0 =1$
}
}
\label{fig:aniso}
\end{figure}

The variation of $\alpha_0$ with $p$ for different $NV_0$ is shown in 
the $log_{10}\alpha_0 - log_{10}p$ plot in figure (\ref{fig:alpha}). 
We see from (\ref{noaniso}), for isotropic inflation,
$\alpha_0$ depends linearly on $NV_0$. Roughly, this continues to happen
even in the presence of anisotropy. From the figure, we note that
the dependence of $\alpha_0$ on $p$ increases substantially for large $p$. 
A good measure of anisotropy is given by the quantity 
${\dot{\sigma}}/{\dot{\alpha}}$ which is equal to 
${{\sigma_0}}/{{\alpha_0}}$ is our case. Figure (\ref{fig:aniso})
shows $log_{10}(\frac{\sigma_0}{\alpha_0}) - log_{10}p$
plots for various $NV_0$. Indeed we see that a small amount of anisotropy 
persists during the inflation. 
Interestingly, anisotropy picks at a certain value of $p$. 
This value, in turn, depends on $NV_0$.

%
%
%

Having found the anisotropic expanding geometry, we now provide a
discussion on the validity of the assumption made  in deriving the 
equations of motion (\ref{eom}). While writing down these equations, we 
expanded the determinant of the action and kept only leading terms in 
derivatives, both for the gauge field and for the scalar. While for
the gauge field, this is perhaps justified as it is $\alpha^\prime$ 
suppressed. However, the reason for throwing away the higher derivative 
terms
involving scalar requires some arguments. This is what we provide below.

Consider the Lagrangian for the matter part.
\begin{equation}
 L = - N V(T) {\sqrt{{\rm det}~[\delta^\mu{}_\nu +
h(T) F^{\mu}{}_\nu + \partial^\mu T \partial_\nu T]}}.
\nonumber
\end{equation}
Now the solution put in, it becomes
\begin{equation}
 \label{Lsoln}
 L= -N \frac{V_0}{T_0^2 t^2} \sqrt{1-\frac{p^2 T_0^4}{V_0^2}-T_0^2} \;.
\end{equation}
Therefore expanding square-root is valid when both $\frac{p T_0^2}{V_0}$
and $T_0$ are small compared to 1. After redefinition of our variable 
as $p \to p V_0$, we have $p T_0^2$ and $T_0$ which should be small.
In the range of our parameters taken in above solution these conditions
are satisfied. For example if we take the first case, that is $N=10^{10}$
and plot $\log_{10}(p T_0^2)$ vs $\log_{10}(p)$, we can see $T_0$ remains
always small compared to 1.
\begin{figure}[H]
 \centering
 \includegraphics[width=0.5\textwidth]{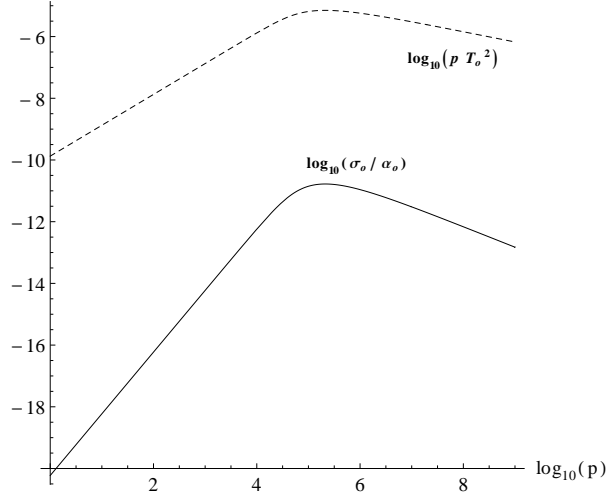}

 \caption{ 
{\em Here we plot both $\frac{\sigma_0}{\alpha_0}$  
and $p\, T_0^2$ (dotted and solid lines respectively). 
One can see that the maximum value $p T_0^2$ reaches is of the order
$10^{-5}$, while $T_0^2$ reaches $10^{-10}$ at the most.
}
}
\label{fig:T0}
\end{figure}

\begin{figure}[H]
 \centering

 \subfloat[]{ 
 \includegraphics[width=0.45\textwidth]{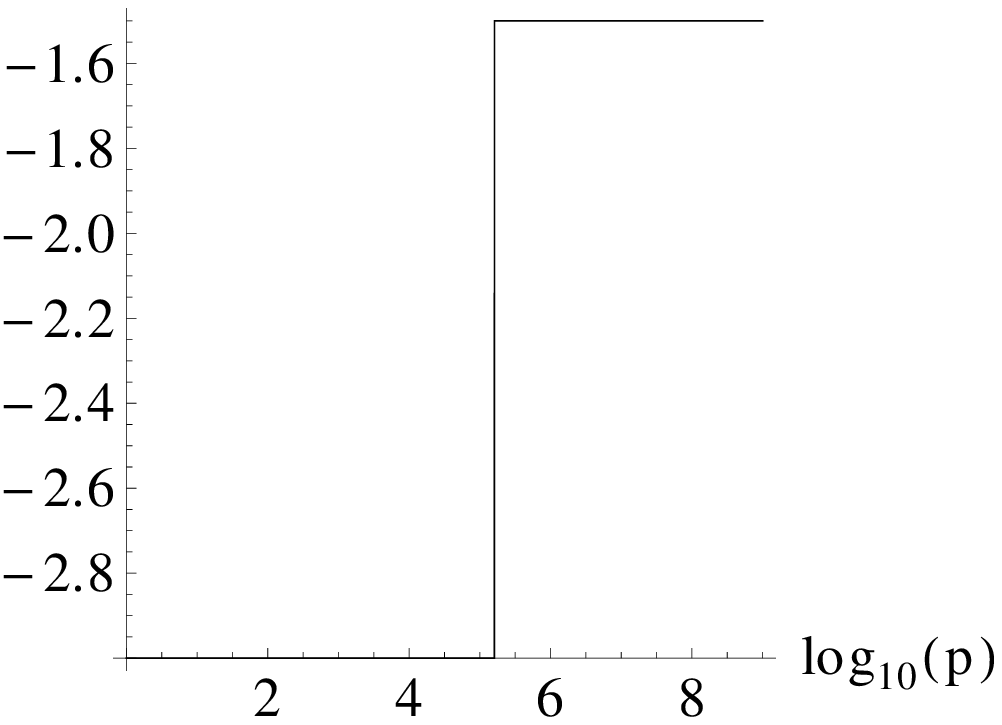}} ~~~~
 \subfloat[]{ 
 \includegraphics[width=0.45\textwidth]{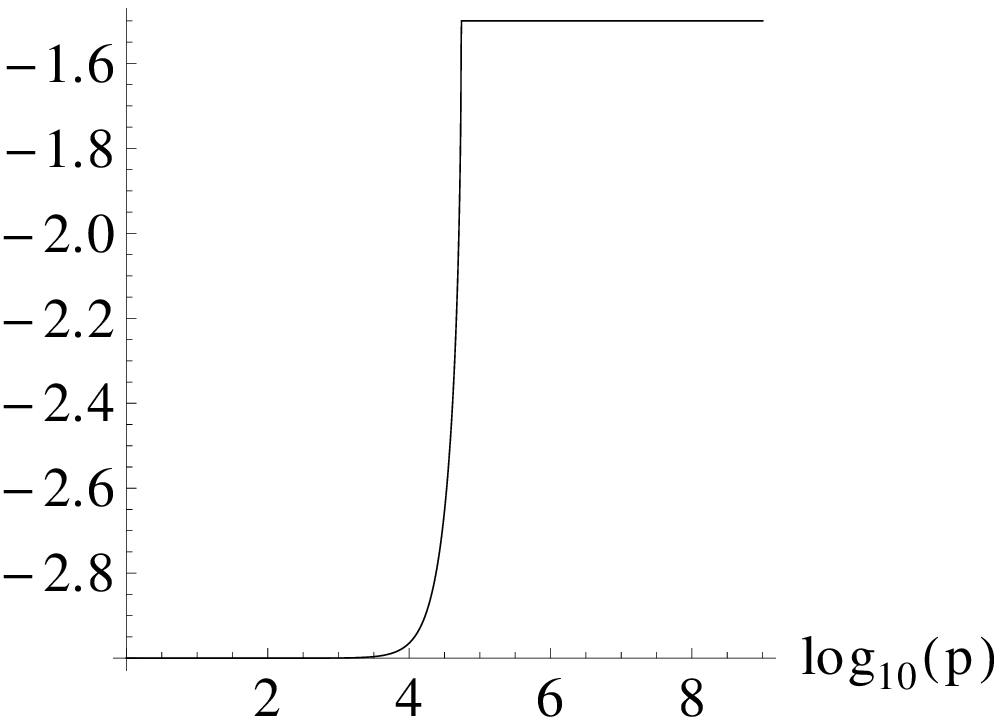}}

 \subfloat[]{
 \includegraphics[width=0.45\textwidth]{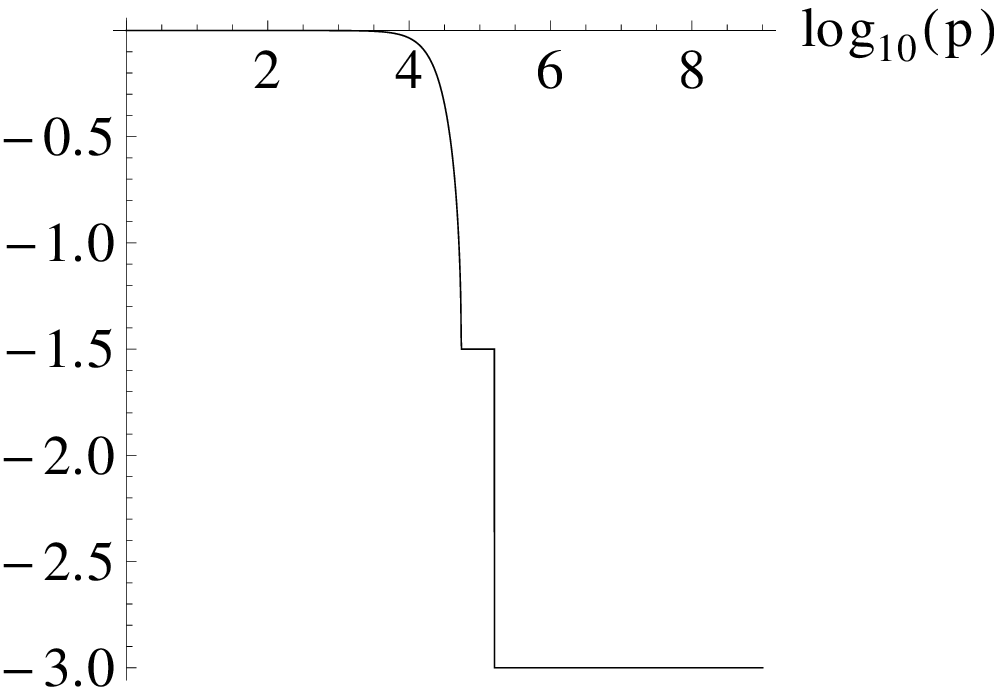}}

 \caption{ 
{\em In these figures, real part of the eigenvalues of $\Lambda$ vs $\log(p)$ 
has been ploted for $N=10^{10}$, with $V_0 = 1$. {\bf \em Figure} ({\bf a}), ({\bf b})
and ({\bf c})
are plots of three different eigenvalues. The graphs are actually smooth, the sharp points 
appear in the graph are because of high rate of change. 
}
}
\label{fig:nu}
\end{figure}

It is important to check the stability of the proposed anisotropic solution
and we do it now. For that, we re-write  the equations of motion (\ref{eom})
in terms new variables
\begin{equation}
X = \frac{\partial_t \sigma}{\partial_t \alpha}, ~~Y = 
{\sqrt{NV}}\frac{\partial_t T}{\partial_t \alpha}, ~~Z = {\sqrt{NU}}e^{- \alpha + 2 
\sigma} \frac{\partial_t A}{\partial_t \alpha}.
\end{equation}
It then follows that 
\begin{eqnarray}
\frac{dX}{d\alpha} &=& X \Big(3 (X^2 -1) + \frac{Y^2}{2} + \frac{Z^2}{3}\Big) + \frac{Z^2}{3},\nonumber\\
\frac{dY}{d\alpha} &=& - \frac{Y^2}{\sqrt{NV_0}} - \frac{2}{\sqrt{NV_0}} \Big(3 (X^2 - 1) +Z^2\Big)
+ Y \Big(3 (X^2 -1) + \frac{Y^2}{2} + \frac{Z^2}{3}\Big) + \frac{Z^2 \partial_T h}{{\sqrt{NV}}h},\nonumber\\
\frac{dZ}{d\alpha} &=& Z \Big( -2 - 2 X + \frac{Y^2}{2} + \frac{Z^2}{3} + 3 X^2  -\frac{Y \partial_T h}{{\sqrt{NV}}h}
+ \frac{Y}{\sqrt{NV_0}}\Big)
\label{rw}
\end{eqnarray}
In writing down the above set of equations, we have made extensive use of (\ref{eom}). Further, in terms of 
$X, Y$ and $Z$, the equations of motion (\ref{eom}) can be re-expressed simply as
\begin{equation}
\frac{dX}{d\alpha} = \frac{dY}{d\alpha} = \frac{dZ}{d\alpha} = 0,
\label{sdle}
\end{equation}
along with the energy conservation condition given in the first equation of (\ref{eom})
\begin{equation}
- \frac{NV}{(\partial_t \alpha)^2} = 3 (X^2 -1) + \frac{Y^2}{2} + \frac{Z^2}{2}.
\end{equation}
It can easily be checked that the anisotropic solution, constructed before, represents a point in (X, Y, Z)
space 
\begin{equation}
X_0 = \frac{\sigma_0}{\alpha_0}, ~~Y_0 = \frac{\sqrt{NV_0}}{\alpha_0}, ~~Z_0 = \frac{\sqrt{NV_0} T_0 p}{\alpha_0}.
\end{equation}
In writing down the above equations we have made the substitution $p \rightarrow p V_0$ as before.
We  now wish to check the nature of the  variations of coordinates around this point. 
Variations of (\ref{rw}), to linear order, around $(X_0, Y_0, Z_0)$
gives
\begin{eqnarray}
\frac{d(\delta X)}{d\alpha} &=& ( -3 + 9 X_0^2 + \frac{Y_0^2}{2} + \frac{Z_0^2}{3}\Big) \delta X + 
X_0 Y_0 \delta Y + \frac{2 Z_0}{3} \Big(1 + X_0\Big) \delta Z, \nonumber\\
 \frac{d(\delta Y)}{d\alpha} &=& 6 X_0 \Big( - \frac{2}{\sqrt{NV_0}} + Y_0\Big) \delta X + \Big(3 (X_0^2 -1) + \frac{3 Y_0^2}{2} + 
\frac{Z_0^2}{3} - \frac{2 Y_0}{\sqrt{NV_0}} \Big) \delta Y\nonumber\\
&&~~~~~ + 2 Z_0 \Big( \frac{\delta}{\sqrt{NV_0}} - \frac{{2}}{\sqrt{NV_0}} + \frac{Y_0}{3}\Big) \delta Z,\nonumber\\
\frac{d(\delta Z)}{d\alpha} &=& 2 Z_0 \Big(3 X_0 -1\Big) \delta X + Z_0 \Big( - \frac{\delta}{\sqrt{NV_0}} +
\frac{1}{\sqrt{NV_0}} + Y_0 \Big) \delta Y\nonumber\\
&&~~~~~ + \Big(-2 - 2 X_0 + 3 X_0^2 - \frac{Y_0 \delta}{\sqrt{NV_0}} + \frac{Y_0}{\sqrt{NV_0}} + \frac{Y_0^2}{2} +  Z_0^2\Big) \delta Z.
\end{eqnarray}
This has the form $A^\prime = \Lambda A$ where $A$ is a column matrix with entries $(\delta X, \delta Y, 
\delta Z)$ and
$\Lambda$ is a three by three matrix with entries written in terms of $(X_0, Y_0, Z_0)$ and $NV_0, 
\delta$. The prime on $A$ denotes derivative with respect to $\alpha$.
Our job is now to simply diagonalize $\Lambda$ and study the nature of its eigen values. These eigen 
values can be computed and expressed solely in terms of $N V_0, p$. However, expressions are very 
messy.
Nevertheless, their natures can be understood from various plots. 
The real part of the three eigen values are plotted in figure (\ref{fig:nu}). The plots 
are for $N=10^{10}$, with $V_0 = 1$. 
Clearly one can see for certain range of $p$ one of the eigenvalues is all most zero and
then it becomes more negative. Real part of other two remain negative for given range of $p$.
Therefore, we conclude that our solution is perturbatively stable.

To summarize, we constructed an anisotropic power law inflationary solution in models where
inflation is achieved by rolling tachyon. We further showed that the solution was perturbatively
stable. Though simple tachyon driven inflationary model suffers from some observational
problems, existence of stable anisotropic inflationary solution is encouraging. Existance of such
solutions also  indicates that  the cosmic no heir conjecture \cite{Wald:1983ky} may require 
appropriate modifications.

\noindent{\bf{Acknowledgments:}} We would like to thank S. Kalyana Rama 
for various comments and
suggestions on a previous version of the paper and Jiro Soda for helpful communications.

\end{document}